\documentclass[epj,nopacs]{svjour}

\usepackage{graphicx}
\usepackage{maybemath}
\usepackage{dcolumn}
\usepackage{latexsym}
\usepackage{indentfirst}
\usepackage{amsmath}
\usepackage{amssymb}
\usepackage{cite}
\usepackage{color}
\usepackage{graphicx}
\usepackage{bm}
\usepackage{maybemath}
\usepackage{widetext}

\newcolumntype{.}{D{x}{}{6}}

\newcommand{\vp}{\mathrm{vp}}

\newcommand{\meV}{\mathrm{meV}}

\newcommand{\muH}{$\mu$H{}}
\newcommand{\muD}{$\mu$D{}}
\newcommand{\muHeThree}{$\mu\,{}^3$He$^+$}
\newcommand{\muHeFour}{$\mu\,{}^4$He$^+$}

\newcommand{\dd}{\mathrm{d}}
\newcommand{\ee}{\mathrm{e}}
\newcommand{\ii}{\mathrm{i}}

\newcommand{\Ei}{\mathrm{Ei}}
\newcommand{\Li}{\mathrm{Li}}

\newcommand{\calO}{\mathcal{O}}

\journalname{EPJ D}

\begin{document}

\title{Semi--Analytic Approach to Higher--Order Corrections
in Simple Muonic Bound Systems:
Vacuum Polarization, Self--Energy and Radiative--Recoil}

\titlerunning{Radiative--Recoil Correction}

\authorrunning{Jentschura and Wundt}

\author{U. D. Jentschura \inst{1,2} \and B. J. Wundt \inst{1}}

\institute{Department of Physics, Missouri University of Science
and Technology, Rolla MO65409, USA \and
Institut f\"ur Theoretische Physik,
Universit\"{a}t Heidelberg,
Philosophenweg 16, 69120 Heidelberg, Germany}

\date{Received: 12--JUL--2011}

\abstract{The current discrepancy of theory and experiment observed
recently in muonic hydrogen necessitates a reinvestigation
of all corrections to contribute to the Lamb shift
in muonic hydrogen (\muH{}), muonic deuterium
(\muD), the muonic ${}^3{\rm He}$ ion (denoted here as \muHeThree), as well as
in the muonic ${}^4{\rm He}$ ion (\muHeFour).
Here, we choose a semi-analytic approach and evaluate
a number of higher-order corrections to vacuum polarization (VP)
semi-analytically, while remaining integrals over the spectral
density of VP are performed numerically.
We obtain semi-analytic results for the second-order correction, and for the
relativistic correction to VP.
The self-energy correction to VP is calculated,
including the perturbations of the Bethe
logarithms by vacuum polarization.  Subleading logarithmic terms in the
radiative-recoil correction to the $2S$--$2P$ Lamb shift of order $\alpha (Z\alpha)^5 \,
\mu^3  \, \ln(Z \alpha)/(m_\mu \, m_N)$ are also obtained.
All calculations are nonperturbative in the mass ratio of
orbiting particle and nucleus.}

\maketitle

%
%
\section{Introduction}
\label{intro}

In simple muonic bound systems
such as muonic hydrogen and deuterium (\muH{} and \muD{}) 
and muonic helium ions (\muHeThree{} and \muHeFour{}),
the mass ratio of the orbiting particle to the mass of the
nucleus is larger than the fine-structure constant $\alpha$. The dominant radiative
correction in these systems is given by electronic vacuum polarization,
which screens the proton charge on a distance scale of the order of the electron Compton
wavelength. Here, by ``electronic''  vacuum polarization,
we refer to the modification of the photon propagator due to the 
creation and annihilation of electron-positron pairs.
Due to the dominance of the vacuum-polarization energy shift
(modification of the Coulomb force law at small distances), 
the $2S$ level is energetically lower in muonic systems as compared
to the $2P_{1/2}$, reversing the ordering found in the hydrogen atom. 

A characteristic property of simple bound muonic systems
is the mass ratio 
\begin{equation}
\label{defxiN}
\xi_N = \frac{m_\mu}{m_N} 
\end{equation}
of orbiting particle and nucleus, which reads as
\begin{subequations}
\label{xi}
\begin{align}
\label{xiH}
\xi_p =& \; \frac{m_\mu}{m_p} = 0.112609\ldots \approx \tfrac{1}{9} \,,
\\[0.77ex]
\label{xiD}
\xi_d = & \; \frac{m_\mu}{m_d} = 0.0563327\ldots \approx \tfrac{1}{18}
\end{align}
for muonic hydrogen and deuterium, and
\begin{align}
\label{xiHe3}
\xi_{\rm He^3} =& \; \frac{m_\mu}{m_{\rm He^3}} = 
0.0376223\ldots \approx \tfrac{1}{26} \,,
\\[0.77ex]
\label{xiHe4}
\xi_{\rm He^4} = & \; \frac{m_\mu}{m_{\rm He^4}} = 
0.0283465\ldots \approx \tfrac{1}{35}
\end{align}
\end{subequations}
for muonic helium ions,
where the latest recommended values of the masses have been
used~\cite{MoTaNe2008}. 
We here denote the masses of the helion nucleus and of the 
alpha particle as $m_{\rm He^3}$ and $m_{\rm He^4}$,
respectively, in contrast to the muonic ions themselves, which we denote 
as ${}^3{\rm He}^+$ and ${}^4{\rm He}^+$.

For heavy muonic ions and atoms, the atomic binding-strength parameter is $Z
\alpha$, where $Z$ is the nuclear charge number and the mass ratio 
$\xi_N$ of
muon and atomic nucleus fulfills 
the inequality 
\begin{equation}
\xi_N = \frac{m_\mu}{m_N} \ll Z\alpha < 1 \,.
\end{equation}
In heavy muonic ions,
 the external-field approximation (Dirac-Coulomb 
equation) gives an excellent result. The
Dirac formalism takes relativistic effects (parameterized by $Z\alpha$) into
account to all orders.  For a system like positronium ($e^+ e^-$) or true
muonium ($\mu^+ \mu^-$, Refs.~\cite{Ma1987,KaJeIvSo1998,BrLe2009}), the
situation is opposite,
\begin{equation}
\alpha = Z\alpha \ll \xi_{\mu^+\mu^-} = \tfrac12 = \calO(1) \,;
\end{equation}
each muon is the ``nucleus for the other one.'' In these systems, the
Breit Hamiltonian is adequate~\cite{BeLiPi1982vol4}.  The (static) Breit
Hamiltonian is exact in the muon-nucleus mass ratio but perturbative in
$Z\alpha$.  For muonic hydrogen,
deuterium and muonic helium ions, the mass ratio
$\xi_N$ is larger than
the atomic binding strength parameter $Z\alpha$.  These systems therefore
``lean'' more toward the situation encountered in positronium and true muonium
than toward heavy muonic ions. It is thus
preferable to treat the reduced-mass dependence of the corrections
exactly (wherever possible).
For completeness, we would like to stress here that \muH{} is the
only muonic atom studied so far 
(Ref.~\cite{PoEtAl2010}) where the mass ratio is larger than the atomic
binding strength parameter. In all other heavy muonic ions
studied primarily in the 1970s and 1980s
(for theoretical overviews see Refs.~\cite{BrMo1978,BoRi1982}), 
the binding parameter $Z\alpha$ is much larger than the mass ratio. 

We proceed as follows.  In Sec.~\ref{relativistic}, we
evaluate second-order as well as relativistic
corrections to vacuum polarization using our semi-analytic approach.
The latter have been the subject of a recent paper~\cite{Je2001pra}.
In Sec.~\ref{selfenergy}, the muon self-energy corrections to vacuum polarization
are evaluated, taking into account the shift of the Bethe logarithms due to
vacuum polarization.  Finally, radiative-recoil corrections are treated in
Sec.~\ref{radrec}, and subleading single logarithmic terms are calculated,
supplementing recent work~\cite{Je2011pra}. Three 
appendices~\ref{appa}---\ref{appc} complement the paper. 
We use natural units with $\hbar = c = \epsilon_0 = 1$.

%
%
\section{Higher--Order Corrections to VP}
\label{relativistic}

\subsection{Second--Order Correction to VP}
\label{secondorder}

For the calculation of VP effects in muonic systems, 
it is convenient to define the ``massive'' Coulomb potential
\begin{equation}
\label{vvp}
v_\vp(\lambda; r) = -\frac{Z\alpha}{r} \, \ee^{-\lambda \, r} \,,
\quad
\lambda \equiv \lambda(\rho) = m_e \, \rho \,,
\end{equation}
where $m_e$ is the electron mass and $\rho$ is a 
dimensionless spectral parameter for the vacuum polarization.
It is also useful to define the linear operator $K$,
\begin{equation}
\label{K}
K[f(\rho)] = \frac{2 \alpha}{3 \pi} \int\limits_2^\infty \dd \rho \;
 \frac{2 + \rho^2 }{\rho^3} \sqrt{1 - \frac{4}{\rho^2}}  \; f(\rho) \,.
\end{equation}
The reduced mass of the system is written as
\begin{equation}
\label{defbetaN}
\mu = \frac{m_\mu \, m_N}{m_\mu + m_N}  \,, \qquad
\beta_N = \frac{m_e}{Z \alpha \, \mu} \,.
\end{equation}
The ratio of the muonic Bohr radius to the
electron Compton wavelength is given by the ratios~\cite{Pa1996mu}
\begin{subequations}
\label{beta}
\begin{align}
\beta_p =& \; 0.7373836\ldots \,, 
\\[0.77ex]
\beta_d =& \; 0.7000861\ldots \,,
\\[0.77ex]
\beta_{\rm He^3} =& \; 0.3438429\ldots \,, 
\\[0.77ex]
\beta_{\rm He^4} =& \; 0.3407691\ldots 
\end{align}
\end{subequations}
Using formulas for the reduced Green function from Appendix~\ref{appb},
and for analytic integrals from Appendix~\ref{appc},
it is possible to analytically evaluate the 
following matrix element, which describes the 
second-order perturbation due to VP.
For the $2P_{1/2}$--$2S_{1/2}$ Lamb shift difference 
\begin{align}
\label{LLres}
& L^{(2)}(\rho_1, \rho_2) = 
\left< 2P \left| v_\vp(\lambda_1; r) \, \frac{1}{(E-H)'} \,
v_\vp(\lambda_2; r) \right| 2P \right> 
\nonumber\\[2ex]
& - \left< 2S \left| v_\vp(\lambda_1; r) \, \frac{1}{(E-H)'} \,
v_\vp(\lambda_2; r) \right| 2S \right> \,,
\end{align}
where the prime denotes the reduced Green function, we find
 
\begin{widetext}
\begin{align}
& L^{(2)}(\rho_1, \rho_2) = (Z\alpha)^2 \mu  \left( 
\frac{ [1+\beta_N (\rho_1 + \rho_2)]^{-5} \, {\mathcal Q}}%
{12 (1+\beta_N \rho_1)^5 (1+\beta_N \rho_2)^5} + 
\frac{\beta_N^2  
\left[ \rho_1^2 + \rho_2^2 + (\beta_N  \rho_1  \rho_2)^2 \right]}
{(1+\beta_N \rho_1)^4  (1+\beta_N \rho_2)^4} 
\ln\left( \frac{(1 + \beta_N  \rho_1)  (1 + \beta_N  \rho_2)}%
{1 + \beta_N  (\rho_1 + \rho_2)} \right) \right) .
\end{align}
Here, ${\mathcal Q} \equiv 
{\mathcal Q}(\beta_N; \rho_1, \rho_2)$ is a polynomial in the 
three arguments, symmetric in $\rho_1$ and $\rho_2$ and rather compact,
\begin{align}
\label{Qres}
{\mathcal Q} =& \;
- 3  \rho_1^2 
+ 18  \beta_N^2  \rho_1^4 
+ 24  \beta_N^3  \rho_1^5 
+ 9  \beta_N^4  \rho_1^6 
- 24  \beta_N  \rho_1^2  \rho_2
- 3  \beta_N^2  \rho_1^3  \rho_2
+ 99  \beta_N^3  \rho_1^4  \rho_2
+ 111  \beta_N^4  \rho_1^5  \rho_2
+ 33  \beta_N^5  \rho_1^6  \rho_2
\nonumber\\[2ex]
& \; 
- 75 \beta_N^2  \rho_1^2   \rho_2^2
- 189  \beta_N^3  \rho_1^3  \rho_2^2
- 33  \beta_N^4  \rho_1^4 \rho_2^2
+ 39  \beta_N^5  \rho_1^5  \rho_2^2
+ 12   \beta_N^6  \rho_1^6  \rho_2^2
- 142  \beta_N^4  \rho_1^3  \rho_2^3
- 62  \beta_N^5  \rho_1^4  \rho_2^3
\nonumber\\[2ex]
& \; 
+ 84  \beta_N^6  \rho_1^5   \rho_2^3
+ 36  \beta_N^7  \rho_1^6  \rho_2^3
+ 74  \beta_N^6  \rho_1^4 \rho_2^4
+ 120  \beta_N^7  \rho_1^5  \rho_2^4
+ 12   \beta_N^8  \rho_1^6  \rho_2^4
- 12  \beta_N^8  \rho_1^5  \rho_2^5
+ (\rho_1 \leftrightarrow \rho_2) .
\end{align}
\end{widetext}
The energy shift is given as 
\begin{equation}
\Delta E^{(2)} = K_1\left[ K_2 \left[ L^{(2)}(\rho_1, \rho_2) \right] \right] \,,
\end{equation}
where $K_1$ and $K_2$ are the generalizations of the operator
in Eqs.~\eqref{K} to integration variables $\rho_1$ and $\rho_2$.
Our semi-analytic approach allows us to evaluate 
the remaining two-dimensional numerical integrals
to essentially arbitrary accuracy.

For the muonic systems of interest,
the second-order shift is found as
\begin{subequations}
\label{DeltaE2vp}
\begin{align}
\Delta E^{(2)}(\mbox{\muH}) = & \; 0.150897 \, \meV \,,
\\[2ex]
\Delta E^{(2)}(\mbox{\muD}) = & \; 0.172023 \, \meV \,,
\\[2ex]
\Delta E^{(2)}(\mbox{\muHeThree}) = & \; 1.677290 \, \meV \,,
\\[2ex]
\Delta E^{(2)}(\mbox{\muHeFour}) = & \; 1.707588 \, \meV \,.
\end{align}
\end{subequations}
For muonic hydrogen, we confirm the entry in Eq.~(28)
of Ref.~\cite{Pa1996mu},
and for \muHeFour, we confirm the results given in 
Eq.~(38) and~(39) of Ref.~\cite{Ma2007}.
Our semi-analytic approach
eliminates any conceivable numerical inaccuracy 
as the reason for (part of) the experimental-theoretical 
disagreement~\cite{PoEtAl2010}.

\begin{figure}[t!]
\includegraphics[width=0.8\linewidth]{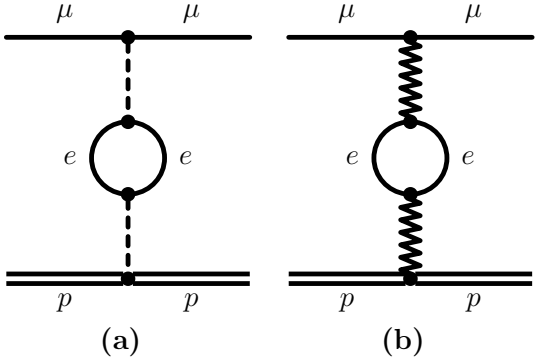}
\caption{\label{fig1} Feynman diagrams for the
relativistic correction to vacuum polarization in the
two-body system of a muon and a nucleus.
The electron-positron
pair in the loop is denoted by the symbol $e$.
Diagram (a) is the Coulomb photon exchange,
given by the $00$-component of the ``massive'' photon propagator,
whereas diagram (b) is the magnetic exchange,
corresponding to the spatial components of the
``massive'' photon propagator.}
\end{figure}

\begin{figure}[t]
\includegraphics[width=1.0\linewidth]{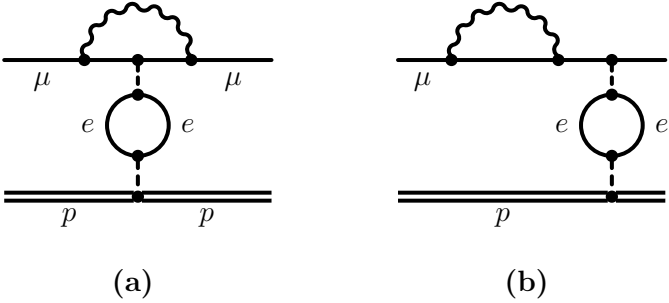}
\caption{\label{fig2} Feynman diagrams for the
self-energy correction to vacuum polarization in a
simple muonic bound system. Diagram~(a) represents the 
vacuum-polarization insertion into the 
exchanged photon in the vertex correction, whereas
diagram (b) represents the wave function correction
to the self energy due to vacuum polarization.}
\end{figure}

%
%
\subsection{Relativistic Correction to VP}
\label{relcorr}

The relativistic correction 
$\delta w =\sum_{i=1}^4 \delta w_i$ to the 
Breit interaction, relevant for the 
$2P_{1/2}$--$2S_{1/2}$ Lamb shift, is given as
the sum of four terms~\cite{Pa1996mu,Je2011pra}
\label{deltaw}
\begin{align}
\delta w_1 =& \; 
\frac{Z\alpha}{8} \left( \frac{1}{m_\mu^2} + \frac{\delta_I}{m_N^2} \right) \,
\left( 4 \pi \delta^3(r) - \frac{\lambda^2 }{r} \, \ee^{-\lambda r} \right) \,,
\\[2ex]
\delta w_2 =& \; -\frac{Z\alpha \lambda^2 \ee^{-\lambda r}}{4 m_\mu m_N r}\,
\left( 1 - \frac{\lambda\,r}{2} \right) \,,
\nonumber\\[2ex]
\delta w_3 = & \; 
-\frac{Z\alpha \; \ee^{-\lambda r} }{4 m_\mu m_N} \;
p^i \; \left( \frac{\delta^{ij}}{r} + \frac{1 + \lambda\, r}{r^3}  \, r^i r^j
\right) \; p^j \,,
\nonumber\\[2ex]
\delta w_4 =& \; 
Z\alpha \left( \frac{1}{4 m_\mu^2} + \frac{1}{2 m_\mu m_N} \right)
\frac{\ee^{-\lambda r} \, (1 + \lambda r)}{r^3} \; \vec \sigma \cdot \vec L \,,
\nonumber
\end{align}
where $\delta_I$ is one for half-integer nuclear 
spin and zero for integer nuclear spin.
For the relevant Feynman diagrams, see Fig.~\ref{fig1}.
The massive Breit Hamiltonian~\eqref{deltaw} is a generalization 
of the ordinary Breit Hamiltonian briefly revisited in Appendix~\ref{appa}.
The relativistic correction to VP is the sum 
of first-order and second-order perturbations,
\begin{subequations}
\label{deltaEEvp}
\begin{align}
\label{deltaEvp}
\delta E_\vp = & \; 
\delta E^{(1)} + \delta E^{(2)} \,,
\\[0.11ex]
\label{EE1}
\delta M^{(1)} = & \;
\left< n \ell_j \left| \delta w \right| n \ell_j \right>  \,,
\\[0.11ex]
\label{EE2}
\delta M^{(2)} = & \;
2 \, 
\left< n \ell_j \left| \delta H
\left( \frac{1}{E_{n\ell} - H} \right)' \, v_\vp \, 
\right| n \ell_j \right> \,,
\\[0.11ex]
\delta E^{(1)} = & \; K [ \delta M^{(1)} ] \,,\qquad
\delta E^{(2)} = K [ \delta M^{(2)} ] \,,
\\[0.11ex]
\delta E_{\rm vp} = & \;
\delta E^{(1)} + \delta E^{(2)} \,.
\end{align}
\end{subequations}
Here, the reference state has
principal quantum number $n$, orbital angular momentum $\ell$,
and total angular momentum $j$.
For the second-order matrix elements
$\delta M^{(2)}$, one needs the explicit formulas for 
the reduced Green function given in Appendix~\ref{appb}.
The radial integrals can be performed analytically
using formulas given in Appendix~\ref{appc}.
Finally, the matrix elements $\delta M^{(1)}$ and 
$\delta M^{(2)}$ can be evaluated analytically.

We present explicit results for the 
matrix elements $\delta M^{(1)}$
and $\delta M^{(2)}$ [see Eqs.~\eqref{EE1} and~\eqref{EE2}], 
assuming $\delta_I = 1$, 
for the muonic bound systems under investigation,
in terms of the parameters $Z\alpha$, 
$\xi_N$ [defined in Eq.~\eqref{defxiN}],
and $\beta_N$ [defined in Eq.~\eqref{defbetaN}],
 
\begin{widetext}
\begin{subequations} 
\label{Mres}
\begin{align} 
\delta M^{(1)}(2P_\frac12) =& \;  - \frac{(Z\alpha)^4 \, \mu \,
\left[ 2 + 10 \, \xi_N + 4 \beta_N \rho (2 + 7 \xi_N) + 
3 (\beta_N \, \rho)^2 (3 + 6 \xi_N + \xi_N^2) \right]}%
{96 \, (1 + \xi_N)^2 \, (1 + \beta_N \, \rho)^4} \,,
\\[2ex]
\delta M^{(1)}(2S_\frac12) =& \;  \frac{(Z\alpha)^4 \, \mu \,
\left\{ 2 \left[ 1 + \xi_N^2 - \xi_N  \, (1 + \beta_N \, \rho)^2 \,
(3 + 4 \beta_N \rho) \right] +
\beta_N \rho \, (1 + \xi_N^2) 
\left[ 8 + \beta_N \, \rho (11 + 8 \beta_N \rho)\right] \right\}}
{32 \, (1 + \xi_N)^2 \, (1 + \beta_N \, \rho)^4} \,,
\\[2ex]
\delta M^{(2)}(2P_\frac12) =& \;  \frac{(Z\alpha)^4 \,  \mu \,
\left[ f(\beta_N, \rho, \xi_N) -
8 (\beta_N \rho)^2 \, (2 + \xi_N + \xi_N^2) - 
8 \, (1 + \beta_N \, \rho) \, (3 + 6 \xi_N + 2 \xi_N^2) \,
\ln(1 + \beta_N \, \rho) \right] }
{96 \, (1 + \xi_N)^2 \, (1 + \beta_N \, \rho)^5} \,,
\\[2ex]
f(\beta_N, \rho, \xi_N) =& \;
-13 -23 \, \xi_N -7 \, \xi_N^2 - 
\beta_N \, \rho \, \left( 41 + 67 \, \xi_N + 19 \, \xi_N^2 \right) \,,
\\[2ex]
\delta M^{(2)}(2S_\frac12) =& \;  \frac{(Z\alpha)^4 \,  \mu \,
\left[ g(\beta_N, \rho, \xi_N) - 
8 \, (\beta_N \, \rho)^4 \, (\xi_N - 1)^2 -
8 \, (1 + \xi_N)^2 \, (1 + \beta_N \, \rho) 
\, \left(1 + 2 (\beta_N \, \rho)^2 \right) 
\, \ln(1 + \beta_N \, \rho) \right] }
{32 \, (1 + \xi_N)^2 \, (1 + \beta_N \, \rho)^5} \,,
\\[2ex]
g(\beta_N, \rho , \xi_N) =& \;
-7 -5 \xi_N -7 \xi_N^2 - 
9 \beta_N  \rho  \left( 3 + \xi_N + 3 \xi_N \right) 
- 2 (\beta_N \rho)^2 (15 - 13 \xi_N + 15 \xi_N^2) -
10 (\beta_N \rho)^3 (3 - \xi_N + 3 \xi_N^2).
\end{align}
\end{subequations} 
\end{widetext}
This reduces the calculation of 
$\delta E^{(1)}$ and
$\delta E^{(2)}$ to simple numerical 
integrals over the spectral representation
of the vacuum polarization ($K$ operator).
We have not attempted here to carry out
the remaining integrals over $\rho$ analytically
because the integral representation
allows us to evaluate the relativistic correction
to VP to essentially arbitrary accuracy.

For the $2P_{1/2}$--$2S_{1/2}$ Lamb shift, we indicate the 
difference as $\Delta E_{\rm vp} = 
\delta E_{\rm vp}(2P_{1/2}) - \delta E_{\rm vp}(2S_{1/2})$,
\begin{subequations}
\label{DeltaEvp}
\begin{align}
\label{DeltaEvpmuH}
\Delta E_{\rm vp}(\mbox{\muH}) = & \; 0.018759\, \meV \,,
\\[2ex]
\label{DeltaEvpmuD}
\Delta E_{\rm vp}(\mbox{\muD}) = & \; 0.021781\, \meV \,,
\\[2ex]
\label{DeltaEvpmuHe3}
\Delta E_{\rm vp}(\mbox{\muHeThree}) = & \; 0.509344 \, \meV \,,
\\[2ex]
\label{DeltaEvpmuHe4}
\Delta E_{\rm vp}(\mbox{\muHeFour}) = & \; 0.521104 \, \meV \,.
\end{align}
\end{subequations}
The relativistic 
correction to vacuum polarization is represented by the 
tree-level Feynman diagrams in Fig.~\ref{fig1},
which represent vacuum-polarization 
insertions in the Coulomb and magnetic exchange,
with relativistic wave functions in the in and out states.
Our results are in agreement with those recently 
reported in Ref.~\cite{Je2011pra}.
This concludes our semi-analytic 
rederivation of the relativistic 
corrections to vacuum polarization,
with proper account of the reduced-mass dependence.

%
%
\section{Self--Energy Correction to VP}
\label{selfenergy}

In Sec.~3.2 of Ref.~\cite{Je2011aop1},
the self-energy correction to vacuum polarization has 
been discussed. In the cited reference, a partially
nonperturbative approach has been chosen in order
to evaluate certain matrix elements in a nonperturbative 
framework, using exact wave functions for the 
Schr\"{o}dinger-Coulomb problem, evaluated on 
exact operators that take vacuum polarization into account.
Here, we compare the nonperturbative
to a perturbative treatment of the vacuum-polarization
potential $V_{\rm vp}$. The Uehling potential
is added to the Schr\"{o}dinger Hamiltonian
by the replacement $V \to V + V_{\rm vp}$.
The effect of high-energy virtual photons
in the self-energy loops given in Fig.~\ref{fig2}
can thus be expressed in terms of the Dirac $F_1$ 
form factor acting on the vacuum polarization
potential $V_{\rm vp}$. 
When rewritten in terms of the noncovariant photon energy cutoff
$\epsilon$, which is a convenient overlapping parameter 
in Lamb shift calculations~\cite{Pa1993}, we have
in first order in $V_{\rm vp}$,
\begin{align}
\delta E_H =& \; \frac{\alpha}{3 \pi m_\mu^2} \, 
\left[ \ln\left( \frac{m_\mu}{2 \epsilon} \right) + \frac{10}{9} \right] \,
\Biggl( \left< n \ell_j \left| \vec\nabla^2 V_{\rm vp} \right| n \ell_j\right>  
\nonumber\\[2ex]
& \; + 
2 \, \left< n \ell_j \left| V_{\rm vp} \left( \frac{1}{E - H} \right)' 
\vec\nabla^2 V \right| n \ell_j \right> \Biggr) \,.
\end{align}
This is equivalent to expanding Eq.~(3.7) of Ref.~\cite{Je2011aop1} 
to first order in the vacuum--polarization potential.
The correction $\delta E_H $ to the high-energy part 
is expressed in terms of a parameter $V_{61}$,
\begin{align}
\delta E_H =& \; 
= \frac{\alpha^2 \, (Z\alpha)^4 \, \mu^3}{\pi^2 m_\mu^2 \, n^3} \, V_{61} \,
\left\{ \ln\left( \frac{m_\mu}{2 \epsilon} \right) + \frac{10}{9} \right\} \,.
\end{align}
We find, using techniques similar to those employed in
Sec.~\ref{relativistic},
\begin{subequations}
\begin{align}
\label{V61res}
V_{61}(2P_{1/2}; \mbox{\muH}) =& \; -0.02327 \,, 
\\[2ex]
V_{61}(2P_{1/2}; \mbox{\muD}) =& \; -0.02453 \,, 
\\[2ex]
V_{61}(2P_{1/2}; \mbox{\muHeThree}) =& \; -0.04345 \,, 
\\[2ex]
V_{61}(2P_{1/2}; \mbox{\muHeFour}) =& \; -0.04369  \,.
\end{align}
\end{subequations}
For the $2S_{1/2}$ state, the results are
\begin{subequations}
\begin{align}
V_{61}(2S_{1/2}; \mbox{\muH}) =& \; 3.08601 \,, 
\\[2ex]
V_{61}(2S_{1/2}; \mbox{\muD}) =& \; 3.18785 \,, 
\\[2ex]
V_{61}(2S_{1/2}; \mbox{\muHeThree}) =& \; 4.71872 \,, 
\\[2ex]
V_{61}(2S_{1/2}; \mbox{\muHeFour}) =& \; 4.73968 \,.
\end{align}
\end{subequations}
The results for \muH{}
confirm the entries in Eq.~(3.8) of Ref.~\cite{Je2011aop1},
where a nonperturbative approach in the vacuum-polarization 
potential was employed.
The correction due to the anomalous magnetic moment of the electron 
reads as 
\begin{align}
\delta E_M =& \; \frac{\alpha}{4 \pi m_\mu^2} \, 
\Biggl( 
\left< n\ell_j\left| \frac{1}{r} \, \frac{\partial V_{\rm vp}}{\partial r} 
\left( \vec\sigma \cdot \vec L \right) \right| n\ell_j\right> 
\nonumber\\[2ex]
& \; + 2 \, \left< n\ell_j \left| V_{\rm vp} \,
\left( \frac{1}{E - H} \right)' \,
\frac{1}{r} \, \frac{\partial V}{\partial r} \,
\left( \vec\sigma \cdot \vec L \right) \right| n\ell_j\right> 
\Biggr)
\nonumber\\[2ex]
=& \; \frac{\alpha^2 \, (Z\alpha)^4 \, \mu^3}{\pi^2 m_\mu^2 \, n^3} \, M_{60} \,.
\end{align}
The matrix element $M_{60}$ is nonvanishing for $P$ states,
\begin{subequations}
\begin{align}
\label{M60res}
M_{60}(2P_{1/2}; \mbox{\muH}) =& \; -0.04276 \,,
\\[2ex]
M_{60}(2P_{1/2}; \mbox{\muD}) =& \; -0.04656 \,,
\\[2ex]
M_{60}(2P_{1/2}; \mbox{\muHeThree}) =& \; -0.13396 \,,
\\[2ex]
M_{60}(2P_{1/2}; \mbox{\muHeFour}) =& \; -0.13556 \,.
\end{align}
\end{subequations}
For $S$ states, we have $M_{60}(2S_{1/2}) = 0$  in view of 
angular symmetry. For the $2P$ state, we here take the opportunity 
to correct the result given in Eq.~(3.10) of Ref.~\cite{Je2011aop1}.
The correction has negligible influence on the numerical
result reported below in Eq.~\eqref{DeltaEsvp} for muonic hydrogen.

The low-energy part is conveniently expressed as
\begin{align}
\delta E_L =& \; 
\frac{\alpha^2 \, (Z\alpha)^4 \, \mu^3}{\pi^2 m_\mu^2 \, n^3} \,
\left[ V_{61} \, \ln\left( \frac{\epsilon}{(Z\alpha)^2 \mu} \right) 
- \frac43 \, L_{60} \right] \,,
\end{align}
where the $V_{61}$ term leads to a cancellation
of the $\epsilon$ parameter.  The coefficient $L_{60}$ gives the 
modification of the Bethe logarithm due to the 
Uehling potential. Numerically, we find for $2P_{1/2}$, 
\begin{subequations}
\label{L60_2P}
\begin{align}
L_{60}(2P_{1/2}; \mbox{\muH}) =& \;       -0.014559 \,,
\\[2ex]
L_{60}(2P_{1/2}; \mbox{\muD}) =& \;       -0.015446 \,,
\\[2ex]
L_{60}(2P_{1/2}; \mbox{\muHeThree}) =& \; -0.032293 \,,
\\[2ex]
L_{60}(2P_{1/2}; \mbox{\muHeFour}) =& \;  -0.032573 \,,
\end{align}
\end{subequations}
whereas for $2S_{1/2}$,
\begin{subequations}
\label{L60_2S}
\begin{align}
L_{60}(2S_{1/2}; \mbox{\muH}) =& \; 11.176 \,, 
\\[2ex]
L_{60}(2S_{1/2}; \mbox{\muD}) =& \; 11.464 \,, 
\\[2ex]
L_{60}(2S_{1/2}; \mbox{\muHeThree}) =& \; 15.640 \,, 
\\[2ex]
L_{60}(2S_{1/2}; \mbox{\muHeFour}) =& \; 15.696 \,.
\end{align}
\end{subequations}
The total self-energy vacuum-polarization
correction to the $2P$--$2S$ Lamb shift then is 
\begin{align}
& \delta E_{\rm svp} = \delta E_H + \delta E_M + \delta E_L
= \frac{\alpha^2 (Z\alpha)^4 \, \mu^3}{\pi^2 m_\mu^2 \, n^3} \;
\\[2ex]
& \;  \times \left\{
\left[ \ln\left( \frac{m_\mu}{2 (Z\alpha)^2 \mu} \right) + 
\frac{10}{9} \right] \Delta V_{61} + \Delta M_{60} 
- \frac43 \Delta L_{60} \right\} \,,
\nonumber
\end{align}
where
\begin{subequations}
\begin{align}
\Delta V_{61} =& \; V_{61}(2P_{1/2}) - V_{61}(2S_{1/2}) \,,
\\[2ex]
\Delta M_{60} =& \; M_{60}(2P_{1/2}) - M_{60}(2S_{1/2}) \,,
\\[2ex]
\Delta L_{60} =& \; L_{60}(2P_{1/2}) - L_{60}(2S_{1/2}) \,.
\end{align}
\end{subequations}
The final numerical values for the contributions to the 
$2P_{1/2}$--$2S_{1/2}$ Lamb shift are given as
\begin{subequations}
\label{DeltaEsvp}
\begin{align}
\Delta E_{\rm svp}(\mbox{\muH}) = & \; -0.00254 \, \meV \,,
\\[2ex]
\Delta E_{\rm svp}(\mbox{\muD}) = & \; -0.00306 \, \meV \,,
\\[2ex]
\Delta E_{\rm svp}(\mbox{\muHeThree}) = & \; -0.06269 \, \meV \,,
\\[2ex]
\Delta E_{\rm svp}(\mbox{\muHeFour}) = & \; -0.06462 \, \meV \,.
\end{align}
\end{subequations}

\begin{figure}[t]
\includegraphics[width=0.7\linewidth]{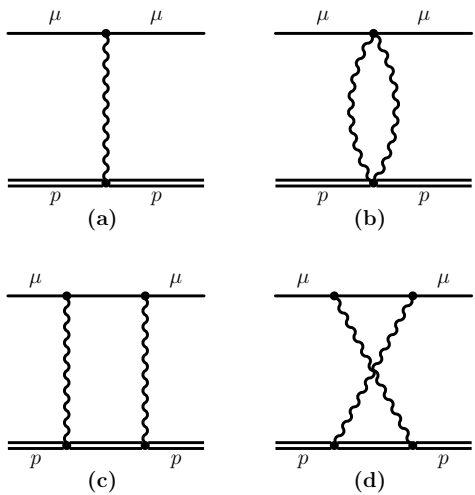}
\caption{\label{fig3} Typical Feynman 
diagrams for the recoil correction.
Diagram (a) is the frequency-dependent part of the Breit
interaction, (b) is the seagull exchange graph,
and (c) and (d) are two-photon exchange graphs.}
\end{figure}

\begin{figure}[t]
\includegraphics[width=0.8\linewidth]{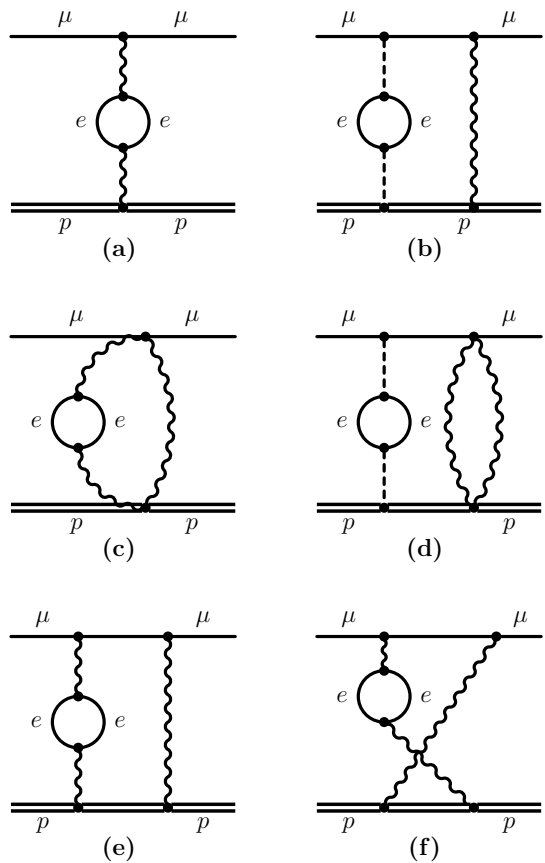}
\caption{\label{fig4} The six Feynman diagrams 
for the vacuum-polarization correction to recoil 
(radiative-recoil correction) of relative 
order $\alpha$, in muonic and antiprotonic systems. 
The electron-positron
pair in the loop is denoted by the symbol $e$.
Coulomb photons are denoted by dashed lines.}
\end{figure}

%
%
\section{Recoil Correction to VP}
\label{radrec}

%
%
\subsection{Formulation of Radiative Recoil}
\label{recoil}

The first genuine two-body energy 
correction beyond the frequency-independent
part of the Breit interaction involves 
a photon-frequency dependent transverse exchange 
and a relativistic two-photon exchange~\cite{Pa1998}.
For the $1S$ state, the corresponding energy 
shift was calculated in Ref.~\cite{Sa1952},
which is why the recoil correction is commonly referred
to as the Salpeter correction, and 
the result was generalized to an arbitrary excited state
in Ref.~\cite{Er1977}.
The correction reads, for an individual 
state with principal quantum number $n$ and angular momentum $\ell$,
\begin{align}
\label{ER}
& E_R =\frac{(Z \alpha)^5 \mu^3}{\pi m_\mu m_N n^3} \, 
\left\{ \frac23 \, \delta_{\ell 0} \ln\left( \frac{1}{Z \alpha} \right) 
- \frac83 \, \ln k_0 - \frac{\delta_{\ell 0}}{9} 
\right.
\nonumber\\[2ex]
& \; \left. 
- \frac{7 a_n}{3}
- \frac{2 \delta_{\ell 0}}{m_\mu^2 - m_N^2} 
\left[ 
m_\mu^2 \ln\left( \frac{m_N}{\mu} \right) -
m_N^2 \ln\left( \frac{m_\mu}{\mu} \right) 
\right] \right\} \,,
\end{align}
where
\begin{equation}
a_n = -2 \left[ \ln \left( \frac{2}{n} \right)  + 
1 + \frac{1}{2 n} + \Psi(n) + \gamma_E \right] \,,
\end{equation}
and $\Psi(n) = \sum_{k=1}^{n-1} k^{-1}$ denotes
the logarithmic derivative of the Gamma function.

For reference, we evaluate the recoil 
correction for the $2P_{1/2}$--$2S_{1/2}$ Lamb shift,
for the muonic systems under investigation,
\begin{subequations}
\label{DeltaER}
\begin{align}
\Delta E_R(\mbox{\muH}) = & \; -0.044971\, \meV \,,
\\[2ex]
\Delta E_R(\mbox{\muD}) = & \; -0.026561\, \meV \,,
\\[2ex]
\Delta E_R(\mbox{\muHeThree}) = & \; -0.558107\, \meV \,,
\\[2ex]
\Delta E_R(\mbox{\muHeFour}) = & \; -0.433032\, \meV \,.
\end{align}
\end{subequations}
We thus confirm the result indicated in Eq.~(66) of 
Ref.~\cite{Ma2007} and the results given in 
Tables 3--6 of Ref.~\cite{Bo2011preprint}.

According to the derivation presented in Ref.~\cite{Pa1998}, the 
recoil correction is formulated in terms 
of four corrections. Two of these are associated
with the photon-frequency dependent part of the 
Breit interaction [see Fig.~\ref{fig3}(a)].
The other two are associated with two-photon exchange
[see Fig.~\ref{fig3}(b), (c) and (d)].
The frequency-dependent one-photon exchange 
in Fig.~\ref{fig3}(a) leads to 
a low-energy part $E_L$ where the
photon frequency fulfills $\omega \ll Z\alpha \, \mu$
and therefore is small against the 
orbiting particle momentum,
and a middle-energy part $E_M$, where 
$\omega \gg Z\alpha \, \mu$.
The two parts are separated by an overlapping parameter. 
The frequency-dependent part of the 
Breit interaction is described by $E_L$ and $E_M$.

For low photon frequencies, the two-photon exchange is described by the seagull
graph in Fig.~\ref{fig3}(b).  The seagull exchange energy correction $E_S$
effectively represents the ``low-energy part'' of the two-photon exchange,
whereas the high-energy part $E_H$ describes a local operator, 
with both virtual photon
momenta and the momenta of the constituent particles being large [see
Figs.~\ref{fig3}(c) and~(d)].  It contributes only for $S$ states.
A second overlapping parameter cancels between
$E_S$ and $E_H$. The seagull exchange diagram in
Fig.~\ref{fig3}(b) exists in nonrelativistic QED (NRQED) 
as opposed to fully relativistic QED, where two-photon
emission out of the same vertex is forbidden.
Indeed, the seagull diagram follows from the two-photon 
exchange diagram of relativistic 
QED (two-photon emission) in the limit of soft photons and 
negative-energy virtual fermion states 
(the fermion lines would have a ``$Z$'' shape in time-ordered 
perturbation theory).

The radiative-recoil correction
of order $\alpha (Z\alpha)^5 \mu^3/(m_\mu \, m_N)$ is the perturbation
of the recoil correction of order $(Z\alpha)^5 \mu^3/(m_\mu \, m_N)$
by an order-$\alpha$ vacuum polarization,
and is calculated here for simple muonic bound systems.
Because the correction is numerically small, 
we restrict our attention to the leading logarithms.
The contributing diagrams at
relative order $\alpha$ are given in Fig.~\ref{fig4}.  One might expect a
seventh diagram to contribute, with a VP correction to a Coulomb photon
exchange in the seagull graph, where the seagull vertices connect to the muon
and proton/deuteron lines on opposite sides of the
Coulomb/Uehling exchange.  However, the seventh diagram only contributes at
relative order $\alpha (Z\alpha)$ and is not considered here.

%
%
\subsection{Logarithm from Middle-- and Low--Energy Part}
\label{log1}

As evident from the derivation in Ref.~\cite{Pa1998}, the 
logarithmic term in Eq.~\eqref{ER} can be broken down as follows,
\begin{align}
\label{ERlog}
& E_R^{\rm log} =\frac{(Z \alpha)^5 \mu^3}{\pi m_\mu m_N n^3} \, 
\left( \frac83 - 2 \right) \, \delta_{\ell 0} \,
\ln\left( \frac{1}{Z\alpha} \right) \,,
\end{align}
where the term with the prefactor ``$8/3$'' comes from the sum of the 
middle- and low-energy parts.
The first logarithmic term in the 
recoil correction (due to 
the middle- and low-energy parts) can thus be expressed 
as a matrix element of the Laplacian of the Coulomb potential $V$,
\begin{equation}
\label{LR}
E_{L} = \frac{2 \, Z \alpha}{3 \pi \, m_\mu \, m_N} \,
\left< n \ell_j \left| \vec \nabla^2 V \right| n \ell_j \right> \, 
\ln\left(\frac{1}{Z \alpha}\right) \,.
\end{equation}
Perturbing the matrix element by the 
vacuum-polarization potential, we obtain the following 
logarithmic ($L$) term in the radiative-recoil correction,
\begin{align}
\label{deltaER}
\delta E_{L} =& \;
\frac{2 Z \alpha \, \ln[(Z\alpha)^{-1}]}{3 \pi \, m_\mu \, m_N} \,
\biggl( \left< n \ell_j \left| \vec \nabla^2 V_\vp \right| n \ell_j \right> 
\nonumber\\[0.77ex]
& \; + 2 \, \left< n \ell_j \left| \vec \nabla^2 V 
\left( \frac{1}{E_{nP} - H_0} \right)' 
V_\vp \right| n \ell_j \right> \biggr)
\nonumber\\[2ex]
=& \; \frac{\alpha}{\pi} \, \xi_N \,
\frac{2 (Z \alpha)^5 \mu^3}{\pi \, m_\mu^2 \, n^3} \, 
V_{61} \, \ln\left( \frac{1}{Z\alpha} \right) \,.
\end{align}
This result is spin-independent and 
in full agreement with Eq.~(5.165) of a recent unpublished 
work (Ref.~\cite{Za2009phd}), where for $P$ states, nonlogarithmic 
terms are calculated as well (we only consider the leading logarithms here).
Results for $V_{61}$ can be found in Eq.~\eqref{V61res}.

For the $2P_{1/2}$--$2S_{1/2}$ Lamb shift, indicated by the 
prefix $\Delta$, the correction evaluates to 
\begin{subequations}
\label{DeltaEL}
\begin{align}
\label{DeltaELmuH}
\Delta E_L(\mbox{\muH}) = & \; -0.000505 \, \meV \,,
\\[2ex]
\label{DeltaELmuD}
\Delta E_L(\mbox{\muD}) = & \; -0.000295\, \meV \,,
\\[2ex]
\label{DeltaELmuHe3}
\Delta E_L(\mbox{\muHeThree}) = & \; -0.005724\, \meV \,,
\\[2ex]
\label{DeltaELmuHe4}
\Delta E_L(\mbox{\muHeFour}) = & \; -0.004431\, \meV \,.
\end{align}
\end{subequations}
This concludes the treatment of the 
logarithmic terms in the radiative-recoil correction
due to the diagrams in Fig.~\ref{fig2}(a) and (b).
It is {\em not} the leading logarithmic term in the 
radiative-recoil correction.

%
%
\subsection{Logarithm from Seagull and High--Energy Part}
\label{log2}

The mechanism for the generation of logarithmic terms
in the seagull part is different from the middle- and low-energy 
parts and cannot be traced to a perturbative modification of the 
matrix element in the term~\eqref{LR}. 
For the first logarithmic contribution,
it is sufficient to consider the seagull exchange diagram
given in Fig.~\ref{fig4}(c);
the high-energy terms from Fig.~\ref{fig4}(e) and~(f)
cancel an intermediate overlapping parameter and 
their contribution is not needed in the evaluation
of the logarithmic terms. The seagull term, with a vacuum polarization
insertion in the exchanged photon, is given as
\begin{align}
& \delta E_S = -\frac{e^4}{2 m_\mu m_N} \,
K \left[ \int 
\frac{\dd^3 k_1}{(2 \pi)^3} 
\frac{\dd^3 k_2}{(2 \pi)^3} 
\frac{1}{\omega_1 \, k_2} \,
\frac{1}{\omega_1 + k_2} \,
\right.
\\[2ex]
& \; \left.
\left( \delta^{ij} - \frac{k^i_1 \, k^j_1}{\omega_1^2} \right) \,
\left( \delta^{ij} - \frac{k^i_2 \, k^j_2}{k_2^2} \right)
\right] \, \left< n\ell_j |
\ee^{\ii \, \left( \vec k_1 + \vec k_2 \right) \cdot \vec r} 
| n \ell_j \right>
\nonumber
\end{align}
where
$\omega_1 = \sqrt{\vec k_1^2 + \lambda^2} = 
\sqrt{\vec k_1^2 + (m_e \, \rho)^2}$
is the photon frequency for massive photons.
From this integral, using a cutoff $\Lambda$ 
as in Eq.~(27) of Ref.~\cite{Pa1998} as an overlapping
parameter that separates the integration region from the 
high-energy part,
we extract the following logarithmic correction terms,
which contribute only for $S$ states,
\begin{align}
\label{deltaES}
\delta E_S =& \;
\frac{4 \alpha \, (Z\alpha)^5 \,\mu^3}{3 \pi^2 m_\mu m_N n^3}
\, \delta_{\ell 0} \,
\\[2ex]
& \; \times \left[ -\ln^2\left(4 Z \alpha \beta_N^2\right) +
\ln(2 \beta_N) \, \ln\left(4 Z \alpha \beta_N^2\right) \right] \, .
\nonumber
\end{align}
The double logarithmic term is formally leading
(we here confirm the result of Ref.~\cite{Je2011pra}).
This correction evaluates to 
\begin{subequations}
\label{DeltaES}
\begin{align}
\Delta E_{S}(\mbox{\muH}) = & \; 0.000414 \, \meV \,,
\\[2ex]
\Delta E_{S}(\mbox{\muD}) = & \; 0.000251 \, \meV \,,
\\[2ex]
\Delta E_{S}(\mbox{\muHeThree}) = & \; 0.006648 \, \meV \,,
\\[2ex]
\Delta E_{S}(\mbox{\muHeFour}) = & \; 0.005174\, \meV \,.
\end{align}
\end{subequations}
There is a second correction for $S$ states.
It is given by the wave function correction
to the leading seagull logarithm
\begin{equation}
\label{SR}
E_S = -\frac{2 (Z\alpha)^2}{m_\mu \, m_p} \,
\left| \phi_{n\ell}(0) \right|^2 \,
\ln\left(\frac{1}{Z\alpha}\right) \,,
\end{equation}
where $\phi_{n\ell}$ is the nonrelativistic
Schr\"{o}dinger wave function.
We note that this term cannot be traced to the 
expectation value of the Laplacian of the 
Coulomb potential, $\langle \vec\nabla^2 V \rangle$,
and therefore the formalism leading to the 
result in Eq.~\eqref{deltaER} is not applicable.
The wave function correction due to the
potential $v_\vp$ is
\begin{equation}
\delta \phi_{nS} =
\left( \frac{1}{E - H} \right)' v_\vp | \phi_{nS} \rangle \,.
\end{equation}
At the origin, the correction can be expressed as 
($2S$ state)
\begin{align}
& \delta \phi_{2S}(0) =
\frac{(Z\alpha \mu)^{5/2}}{4 \sqrt{2 \pi} (\lambda + Z\alpha \mu)^5} \,
\left[ 4 \lambda^4 + 12 (Z\alpha \mu) \lambda^3 \right.
\nonumber\\[2ex]
& \left. 
+ 4 (Z\alpha \mu)^2 \lambda^2  
+ 11 (Z \alpha \mu)^3 \lambda + 3 (\alpha \mu)^4 \right] 
\nonumber\\[2ex]
& \; 
+ \frac{(Z \alpha \mu)^{7/2} \, \left( 2 \lambda^2 + (\alpha \mu)^2 \right) }%
{ \sqrt{2 \pi} (\lambda + Z \alpha \mu)^4 }
\ln\left( 1 + \frac{\lambda}{Z \alpha \, \mu} \right) \,.
\end{align}
The corresponding logarithmic energy correction for an $nS$ state 
due to the wave function correction reads as 
\begin{equation}
\label{deltaEW}
\delta E_W = - \frac{2 \, (Z \alpha)^2}{m_\mu m_N} \,
K\left[ \; 2 \phi_{n\ell}(0) \; \delta\phi_{n\ell}(0) \; \right]
\ln\left( \frac{1}{Z \alpha} \right) 
\end{equation}
for an individual state.
A numerical evaluation for the Lamb shift 
($2P_{1/2}$--$2S_{1/2}$ energy difference) leads to the results
(again denoted by the prefix $\Delta$ instead of $\delta$),
\begin{subequations}
\label{DeltaEW}
\begin{align}
\Delta E_{W}(\mbox{\muH}) = & \; 0.000228 \, \meV \,,
\\[2ex]
\Delta E_{W}(\mbox{\muD}) = & \; 0.000138 \, \meV \,,
\\[2ex]
\Delta E_{W}(\mbox{\muHeThree}) = & \; 0.004017 \, \meV \,,
\\[2ex]
\Delta E_{W}(\mbox{\muHeFour}) = & \; 0.003123 \, \meV \,,
\end{align}
\end{subequations}
which is of the expected magnitude.
The correction $\Delta E_W$ corresponds to the diagram in Fig.~\ref{fig4}(d).
The seagull and high-energy parts do not generate
any logarithmic radiative-recoil corrections for $P$ states.

%
%
\subsection{Total Logarithmic Radiative--Recoil Correction}

For the Lamb shift, the total logarithmic radiative-recoil
correction is given as the sum
\begin{equation}
\Delta E_{RR} = \Delta E_L + \Delta E_S + \Delta E_W \,.
\end{equation}
It evaluates to
\begin{subequations}
\label{DeltaERR}
\begin{align}
\Delta E_{RR}(\mbox{\muH}) = & \; 0.000136 \, \meV \,,
\\[2ex]
\Delta E_{RR}(\mbox{\muD}) = & \; 0.000093\, \meV \,,
\\[2ex]
\Delta E_{RR}(\mbox{\muHeThree}) = & \; 0.004941 \, \meV \,,
\\[2ex]
\Delta E_{RR}(\mbox{\muHeFour}) = & \; 0.003867 \, \meV \,.
\end{align}
\end{subequations}
The results are numerically small and suppressed with respect 
to the leading recoil correction given in Eq.~\eqref{DeltaER} 
by a factor $\alpha$.

There is some numerical cancellation between the 
leading squared logarithm given by the first 
term on the right-hand side of Eq.~\eqref{deltaES}
and the subleading single logarithms.

%
%
\section{Conclusions}
\label{conclu}

The conclusions of this paper are twofold.  The first conclusion is that a
number of nontrivial higher-order corrections to the Lamb shift in simple
muonic systems (\muH, \muD, \muHeThree, \muHeFour) can be evaluated
semi-analytically.  Certain well-defined, remaining one- or two-dimensional
numerical integrals can easily  be evaluated to essentially arbitrary
accuracy.  The second-order VP shift (Sec.~\ref{secondorder}) has
been investigated numerically in a number of previous
works~\cite{Pa1996mu,Ma2007,Je2011aop1}, and we here confirm these results
using our semi-analytic approach [see Eq.~\eqref{LLres}] and complement the
literature with a result for \muHeThree. The second-order correction is
otherwise easy to evaluate numerically; the only advantage of our approach is
that the integral representation easily allows us to evaluate the
correction to essentially arbitrary accuracy, yielding 
an additional confirmation
for the literature values.  The relativistic correction to VP (with a proper
account of the reduced-mass dependence) is analyzed in Sec.~\ref{relcorr}.  The
calculation is reduced to one-dimensional parametric integrals [see
Eqs.~\eqref{deltaEEvp} and~\eqref{Mres}].  These expressions yield additional
evidence for the results given in Eq.~\eqref{DeltaEvp}.  We have not attempted
to find analytic expressions for the remaining one- and two-dimensional
integrations over the spectral parameter $\rho$ of the
vacuum polarization, but note that such calculations may be
possible~\cite{Pu1957,Ka1998muonic}; we leave this problem for future
investigations.

The second point of the paper is the calculation of two
nontrivial higher-order corrections to
vacuum polarization in simple muonic bound systems:
namely, the self-energy and recoil corrections to VP
(see Secs.~\ref{selfenergy} and~\ref{radrec}).
For the self-energy correction to VP,
the final numerical results are given in Eq.~\eqref{DeltaEsvp}.
Numerically, we confirm results reported previously in 
Ref.~\cite{Je2011aop1} for \muH{}.
The same analytic techniques are used as those 
employed in Sec.~\ref{relativistic}
but intermediate results are suppressed.
In Eq.~\eqref{M60res}, we take the opportunity 
to correct~\cite{Je2011aop1} a computational error in the evaluation of the 
self-energy correction for the matrix element 
$M_{60}$ which describes the vacuum-polarization correction
to the electron anomalous magnetic moment term in the 
Lamb shift (the correction has negligible effect on the 
numerical value of the final result).
Our results include the correction to the Bethe logarithm
due to vacuum polarization
[see Eq.~\eqref{L60_2P} and~\eqref{L60_2S}].

The radiative-recoil correction is found to be numerically small for simple
muonic bound systems. Leading and subleading logarithmic 
terms are calculated here. The first logarithmic term 
in the radiative-recoil correction is due to 
the vacuum-polarization correction to the frequency-dependent 
Breit exchange and is given  in Eq.~\eqref{deltaER}
[see Figs.~\ref{fig4}(a) and (b)]. From
the seagull part, the leading logarithm is given (for $S$ states) by the
vacuum-polarization insertion into the exchange photon lines [see
Figs.~\ref{fig4}(c), (e), (f)]. 
The corresponding double-logarithmic term $\delta E_S$
can be expressed analytically, scales as $n^{-3}$ where $n$ is the principal
quantum number, and can be found in Eq.~\eqref{deltaES}.
The appearance of
the parameter $\beta_N$ in the result implies that the logarithmic coefficient
depends on the mass ratio of muon and electron, as it should.  Finally, the
wave function correction to $S$ states which is nonvanishing, generates a third
logarithm, $\delta E_W$, given in Eq.~\eqref{deltaEW} [see also
Fig.~\ref{fig4}(d)]. 
The total results for the radiative-recoil 
correction (logarithmic terms) for the systems under investigation
can be found in Eq.~\eqref{DeltaERR}.

Self-energy corrections to vacuum polarization
and radiative-recoil corrections are found to be 
of minor significance,
in accordance with simple 
order-of-magnitude estimates.
Still, the calculations are necessary in order 
to exclude a conceivable large logarithmic
term in the higher-order effects as an 
explanation for (part of) the observed spectroscopic
discrepancy~\cite{PoEtAl2010}.
Also, the calculation of the subleading logarithmic 
terms in the radiative-recoil correction clarifies 
the size of one of the traditionally most elusive 
corrections for two-body bound systems, in the
case of a bound muon.
Experiments on these systems are 
ongoing at PSI (Paul--Scherrer--Institute, Villigen).

%
%
\section*{Acknowledgments}

Support by NSF and helpful conversations with 
K.~Pachucki are gratefully acknowledged.  The authors have been
supported by a NIST precision measurement grant.

\appendix

%
%
\section{Breit Hamiltonian}
\label{appa}

The Breit Hamiltonian describes the fine- and hyperfine 
splitting in two-body systems in the order $(Z\alpha)^4$, 
exact to all orders in the mass ratio.
For the $2P$--$2S$ Lamb shift, the relevant terms 
in the Breit Hamiltonian read 
\begin{align}
\label{deltaH}
\delta H =& \; \sum_{j=1}^{4} \delta H_j \,,
\qquad
\delta H_1 =  - \frac{\vec p^{\,4}}{8 m_\mu^3 } - \frac{\vec p^{\,4}}{8 m_N^3 }  \,,
\nonumber\\
\delta H_2 =& \;  \left( \frac{1}{m_\mu^2} + \frac{\delta_I}{m_N^2} \right) 
\frac{\pi Z \alpha \, \delta^3(r)}{2} \,,
\nonumber\\
\delta H_3 =& \; -\frac{Z\alpha}{2 m_\mu m_N} 
p^i \left( \frac{1}{r} + 
\frac{r^i \, r^j}{r^3} \right) p^j \,,
\nonumber\\
\delta H_4  =& \;  \frac{Z\alpha}{r^3} \,
\left( \frac{1}{4 m_\mu^2} + \frac{1}{2 m_\mu m_N} \right)
\; \vec \sigma \cdot \vec L \,,
\end{align}
where the summation convention is used.
Here, $\delta_I = 1$ for half-integer
and $\delta_I = 0$ for integer nuclear spin (see~\cite{PaKa1995}).
The expectation values of 
of the Breit Hamiltonian,
evaluated for the $2P_{1/2}$--$2S_{1/2}$ difference read as
\begin{align}
\label{LmuH}
L(2P_{1/2} \!-\! 2S_{1/2}) =& \; 
\left\{ \begin{array}{cc} 
\dfrac{(Z \alpha)^4 \mu^3}{48 \, m_N^2}  & \qquad \delta_I = 1 \,, \\[2.77ex]
\dfrac{(Z \alpha)^4 \mu^3}{12 \, m_N^2}  & \qquad \delta_I = 0 \,,
\end{array}  \right.
\end{align}
where we recall that $2P_{1/2}$ is energetically higher.
The shift evaluates to 
\begin{subequations}
\label{Lres}
\begin{align}
L(\mbox{\muH}) = & \; 0.05747 \, \meV \,,
\\[2ex]
L(\mbox{\muD}) = & \; 0.06722 \, \meV \,,
\\[2ex]
L(\mbox{\muHeThree}) = & \; 0.12654 \, \meV \,,
\\[2ex]
L(\mbox{\muHeFour}) = & \; 0.29518\, \meV \,.
\end{align}
\end{subequations}
These results confirm entries in Ref.~\cite{Je2011pra}.

%
%
\section{Reduced Green Functions}
\label{appb}

We use the reduced Green function $G'$ of 
the $2S$ state for general radial arguments,
\begin{align}
G'_{2S}(\vec r_1, \vec r_2) =& \; 
\left< \vec r_1 \left| \left( \frac{1}{H_0-E_{2S}} \right)' \right| \vec r_2 \right> 
\nonumber\\
=& \; 
R_{2S}(r_1, r_2) \, Y_{00}(\hat r_1) \, Y^*_{00}(\hat r_2) \,,
\end{align}
where $R_{2S}(r_1, r_2)$ has the following representation,
\begin{align}
\label{R2S}
& R_{2S}(r_1, r_2) = -\frac{\alpha \, \mu^2 \, 
\exp\left(-\frac12 \, \left( R_< + R_> \right) \right)}%
{4 \, R_< \, R_>}
\nonumber\\[0.77ex]
& \; \times 
\left[ 4 \, \ee^{R_<} \, R_< \, (R_> - 2) \, R_>  + Q(R_<, R_>) 
\right.
\nonumber\\[0.77ex]
& \; \qquad \left. - 4 \, R_< \, (R_<  - 2) \, R_> (R_> - 2) \,
F(R_<, R_>) \right] \,.
\end{align}
Here, $R_< = \alpha \mu r_<$ and $R_> = \alpha \mu r_>$, 
with $r_< = \min(r_1, r_2)$, and $r_> = \max(r_1, r_2)$.
The result~\eqref{R2S} constitutes a confirmation of 
Eq.~(23) of Ref.~\cite{Pa1996mu}.
The function $F$ is expressed as
\begin{subequations}
\begin{equation}
F(R_<, R_>) = {\rm Ei}(R_<) - 2 \gamma_E - \ln(R_< \; R_>) \,.
\end{equation}
The exponential integral can be written as
\begin{align}
{\rm Ei}(x) =& \; \int_0^x \dd t\, \frac{\ee^t - 1}{t} + \gamma_E + \ln(|x|) 
\nonumber\\
=& \; - ({\rm P.V.}) \int_{-x}^\infty \dd t\, \frac{\ee^{-t}}{t} = -{\rm E}_1(-x) \,,
\end{align}
\end{subequations}
for real $x$ (positive or negative),
where $\gamma_E = 0.577216\ldots$ is the Euler--Mascheroni constant.
We denote the principal value by the symbol ``P.V.''.
The polynomial $Q$ in Eq.~\eqref{R2S} reads
\begin{align}
& Q(R_<, R_>) = - 8 \, (R_< + R_>)
+ 26 \, R_< \, R_> \, (R_< + R_>) 
\nonumber\\
& \; 
+ R_<^2 \, R_>^2 (R_< + R_>) 
+ 4 \left( R_<^2 - 3 R_< R_> + R_>^2 \right) 
\nonumber\\
& \; - R_< \, R_> \left( 2 R_<^2 + 23 R_< R_> + 2 R_>^2 \right) \,.
\end{align}
Using the integrals listed in Appendix~\ref{appa}, it is 
actually possible to calculate the wave function correction
\begin{equation}
\delta \phi_{nS} = 
\left( \frac{1}{E_{nS} - H_0} \right)' v_\vp | \phi_{nS} \rangle 
\end{equation}
analytically,
where $v_\vp$ is given in Eq.~\eqref{vvp}.

%
%
\section{Integrals}
\label{appc}

In order to carry out the radial integrations
in the calculation of the 
wave function perturbation
\begin{equation}
| \delta \psi_{n \ell_j} \rangle = 
\left( \frac{1}{E_0 - H_0} \right)' v_\vp | n \ell_j \rangle
\end{equation}
for $2S$ and $2P$, one needs integrals with 
integration domains $r \in (0,s)$,
and $r\in (s,\infty)$, due to radial ordering.
We give the results for two useful integrals of the first category:
\begin{subequations}
\label{C1}
\begin{align}
& \int_0^s \dd r \, \ee^{-a \, r} \, \ln(r) 
\\[0.77ex]
& \qquad
= \frac{1}{a} \, \left[ \Ei(-a \, s ) - \gamma_E - 
\ln(a) \, \left( \ee^{-a \, s} + 1 \right) \right]  \,, 
\nonumber\\[2ex]
& \int_0^s \dd r \, \ee^{-a \, r} \, \Ei(b \, r) 
\\[0.77ex]
& \;\;\;
= -\frac{1}{a} \, \left[ 
\Ei\left( (b-a) \, s \right) - \ee^{-a \,s} \, \Ei\left( b \, s \right) 
+ \ln\left( \left| \frac{b}{a - b} \right| \right) \right] 
\nonumber
\end{align}
\end{subequations}
In the second category, the following 
expressions are relevant:
\begin{subequations}
\label{C2}
\begin{align}
& \int_s^\infty \dd r \, \ee^{-a \, r} \, \ln(r) 
= \frac{1}{a} \, \left[ \Ei(-a \, s) \, \ln(a) - \Ei(-a\,s) \right] \,,
\\
& \int_s^\infty \dd r \, \ee^{-a \, r} \, \Ei(b \, r) 
= \frac{1}{a} \, \left[ 
\Ei\left( (b-a) \, s \right) - \ee^{-a \,s} \, \Ei\left( b \, s \right) \right] \,,
\end{align}
\end{subequations}
where we note that the results are much more compact than for 
the first category. 
Finally, integrals over the interval $r \in (0,\infty)$ 
are needed for the evaluation of second-order 
matrix elements [see Eq.~\eqref{LLres}]. 
One example is
\begin{align}
\label{C3}
& \int_0^\infty \dd r \, \ee^{-a \, r} \, \Ei(b \, r) \, \ln(r) = 
\frac{\pi^2}{3 \, a} + \frac{1}{2 \, a} \, \ln^2\left[ a (a-b) \right] 
\nonumber\\[0.77ex]
& \qquad
+ \frac{1}{a} \, \ln(a-b) \; \left[ \gamma_E + \ln\left(\frac{a}{b}\right) \right]
\nonumber\\[0.77ex]
& \qquad
- \frac{1}{a} \left[ \gamma_E \, \ln(b) + \Li_2\left(\frac{a-b}{a}\right) \right] \,.
\end{align}

\end{document}